\documentclass[preprint]{aastex}
\begin{document}

\title{Discovery of An Unusually Blue L Dwarf Within 10~pc of the Sun\footnote{This paper includes data gathered with the 6.5 m Magellan Telescopes
located at Las Campanas Observatory, Chile.}}

\author{Sarah J. Schmidt\altaffilmark{1}, Andrew A. West \altaffilmark{2,3}, Adam J. Burgasser \altaffilmark{2,4,5}, John J. Bochanski \altaffilmark{2}, Suzanne L. Hawley \altaffilmark{1}}

\altaffiltext{1} {Department of Astronomy, University of
Washington, P.O. Box 351580, Seattle, WA 98195, USA}
\altaffiltext{2} {Massachusetts Institute of Technology, Kavli Institute for
Astrophysics and Space Research, Building 37, 77 Massachusetts
Avenue, Cambridge, MA 02139, USA}
\altaffiltext{3} {Department of Astronomy, Boston University, 725 Commonwealth Ave, Boston, MA, 02215}
\altaffiltext{4} {Department of Physics, University of California, San Diego, CA 92093, USA}
\altaffiltext{5} {Visiting Astronomer at the Infrared Telescope Facility, which is operated
by the University of Hawaii under Cooperative Agreement no. NCC 5-538 with
the National Aeronautics and Space Administration, Science Mission
Directorate, Planetary Astronomy Program}

\begin{abstract}
We report the discovery of an unusually blue L5 dwarf within 10~pc of the Sun from a search of Sloan Digital Sky Survey (SDSS) spectra. A spectrophotometric distance estimate of  8.0$\pm$1.6~pc places SDSS J141624.08+134826.7 among the six closest known L dwarfs. SDSS 1416+13 was overlooked in infrared color-based searches because of its unusually blue \textit{J$-$K$_S$} color, which also identifies it as the nearest member of the blue L dwarf subclass. We present additional infrared and optical spectroscopy from the IRTF/SpeX and Magellan/MagE spectrographs and determine \textit{UVW} motions that indicate thin disk kinematics. The inclusion of SDSS 1416+13 in the 20~pc sample of L dwarfs increases the number of L5 dwarfs by 20\% suggesting that the L dwarf luminosity function may be far from complete.
\end{abstract}

\keywords{brown dwarfs; stars: individual (SDSS J141624.08+134826.7); stars: low-mass; stars: luminosity function, mass function} 

\section{Introduction}
\label{sec:intro}
Nearby stars are useful for a wide variety of astronomical investigations, including luminosity functions \citep[e.g.][]{Reid2002,Henry2006,Cruz2007} and mapping the local Galactic potential 
\citep[e.g.,][]{Dehnen1998,Fuchs2009}. Characterizing the solar neighborhood is particularly important for the study of L and T dwarfs, as their intrinsically faint luminosities limit detailed investigations to the brightest members of these spectral types. Recent work has emphasized the assembly of a complete sample of L dwarfs within 20~pc \citep{Cruz2003,Cruz2007,Kendall2004,Reid2008}. This volume-complete sample has been used to determine the luminosity function \citep{Cruz2007} and constrain the mass function across the hydrogen-burning minimum mass limit \citep[e.g.][]{Allen2005}. 

Detailed investigation of the L dwarfs has revealed a subclass of blue L dwarfs, approximately 15 objects with unusually blue \textit{J $-$ K$_S$} colors compared to their spectral types \citep{Cruz2003,Knapp2004,Chiu2006,Cruz2007,Burgasser2008a}. Their near-infrared spectra exhibit stronger FeH, K I and H$_2$O features compared to normal L dwarfs of the same optical spectral type. The spectral slope that produces the blue \textit{J $-$ K$_S$} color is likely due to reduced condensate absorption at 1$~\micron$, possibly the result of a thin or patchy cloud deck \citep{Ackerman2001,Knapp2004,Burgasser2008a}. These cloud peculiarities may be explained by higher surface gravity in the blue L dwarf atmospheres, a consequence of older ages \citep{Folkes2007,Burgasser2008a}, as supported by the lack of Li I detections in their optical spectra \citep{Burgasser2008a}, and a large dispersion in their tangential velocities, compared to L dwarfs with normal colors \citep{Faherty2009}. The cloud decks of these blue L dwarfs could also be modulated by rotation, viewing orientation, or other atmospheric processes.

The Sloan Digital Sky Survey \citep[SDSS;][]{Abazajian2009} has been an invaluable tool for finding L dwarfs \citep{Hawley2002,West2004,Knapp2004,Chiu2006,Schmidt2009} and has been particularly fruitful in identifying blue L dwarfs. While blue L dwarfs have \textit{J$-$K$_S$} colors indistinguishable from those of the brighter late-M dwarfs, a significant contaminant in near-infrared color-selected searches \citep[e.g.][]{Cruz2003}, their optical colors are similar to those of typical L dwarfs. In this paper, we report the discovery of a bright L dwarf in the SDSS data, which is both within 10~pc of the Sun and unusually blue for its spectral type. Section 2 presents SDSS and Two Micron All Sky Survey (2MASS) photometry, spectroscopic observations, and kinematics of this source. Section 3 discusses SDSS 1416+13 in the context of other blue L dwarfs and the low-mass luminosity function.

\section{Observations}
\subsection{SDSS and 2MASS Data}
SDSS J141624.08+134826.7 (hereafter SDSS 1416+13) was discovered in a search of the SDSS seventh data release \citep[DR7;][]{Abazajian2009} as part of an effort to expand the L dwarf census \citep{Schmidt2009}. We selected a sample of 13,629 objects with SDSS spectra and red colors \citep[$i-z$ $>$ 1.4;][]{West2008} and assigned each object a spectral type using the HAMMER spectral typing facility \citep{Covey2007}. Each spectrum was then visually compared to spectral standards. Examination of this large number of spectra allowed us to find L dwarfs without using the restrictive photometric criteria that are needed to exclude the large numbers of M dwarfs with similar colors.

\begin{figure}
\includegraphics[width=0.97\linewidth]{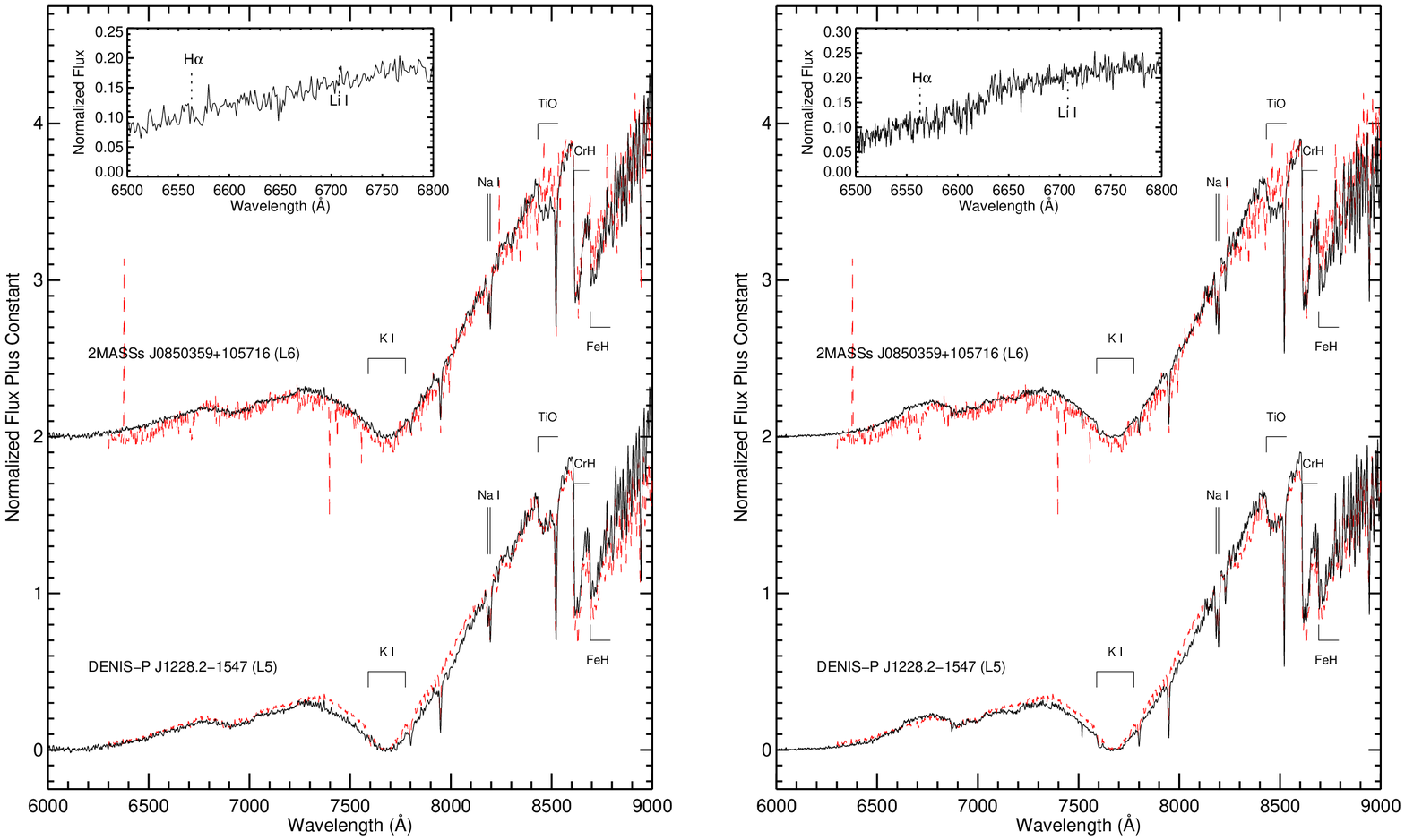} 
\caption{The SDSS spectrum (left panel) and MagE spectrum (right panel) of SDSS 1416+13 (black) shown with optical spectral standards (red) DENIS-P J1228.2-1547 (L5) and 2MASSs J0850359+105716 (L6) from \citet{Kirkpatrick1999}. The MagE data are not corrected for telluric absorption. All spectra are normalized at 8200\AA~and important spectral features are labeled. The spectrum of SDSS 1416+13 from 6500 - 6800\AA~is shown in the inset box, highlighting the region around the H$\alpha$ and Li I lines.} \label{fig:oST_mg}
\end{figure}

Out of the initial 13,629 candidates, we identified a total of 484 L dwarfs in SDSS DR7. For each of these objects, we compiled SDSS photometry and cross-matched to 2MASS \citep{2MASS} with a  search radius of 5$\arcsec$. SDSS 1416+13 was identified as a bright mid-type L dwarf based on its SDSS spectrum and tagged as particularly interesting due to its brightness and apparent proximity, as well as its blue \textit{J$-$K$_S$} colors. SDSS and 2MASS magnitudes, colors, and kinematics for SDSS 1416+13 are given in Table~\ref{tab:prop} and the SDSS spectrum is shown in Figure~\ref{fig:oST_mg}.

\subsection{Optical Spectroscopy}
SDSS 1416+13 was observed with the Magellan Echellette Spectrograph \cite[MagE;][]{Marshall2008} on 2009 May 31 (UT).  MagE is a medium--resolution ($R \sim 4100$) optical spectrograph, with coverage from $\sim$ 3000 to 10,000 \AA.  SDSS 1416+13 was observed in two consecutive exposures of 2100s each.  The 0$^{\prime \prime}$.85 slit was employed for each observation, and ThAr arcs were obtained after the science observation.  The data were reduced using the MagE Spectral Extractor (MASE) pipeline \citep{Bochanski2009}, a comprehensive suite of IDL utilities that flat field, wavelength calibrate, extract, and flux calibrate each spectrum.  The co-added spectrum of SDSS 1416+13 is shown compared to optical spectral standards in Figure \ref{fig:oST_mg}.

Visual comparison of the MagE spectrum to the \citet{Kirkpatrick1999} spectral standards yields a spectral type of L5 for SDSS 1416+13. No Li I ($\lambda$ = 6708\AA) absorption or H$\alpha$ ($\lambda$ = 6563\AA) emission was detected, with upper limits of H$\alpha$ equivalent width (EW) $>$-0.04\AA~and Li I EW $<$ 0.3\AA. H$\alpha$ is uncommon in late-L dwarfs \citep{Gizis2000,Kirkpatrick2000,Schmidt2007} due to the difficulty in producing H$\alpha$ in their mostly neutral atmospheres \citep{Mohanty2002}. The lack of Li I detected in other L5 dwarfs \citep[e.g.,][]{Kirkpatrick2000} places SDSS 1416+13 above the minimum mass for lithium burning \citep[0.06M$_{\odot}$;][]{Burrows2001}. Assuming T$_{eff}$ = 1722~K for an L5 dwarf \citep{Looper2008a} and the evolutionary models of \citet{Burrows1997}, this mass limit corresponds to a lower age limit of $>$0.8~Gyr.

We estimated the distance to SDSS 1416+13 using the M$_i$ versus $i-z$ and M$_i$ versus $i-J$ relations derived in \citet{Schmidt2009}:

\begin{equation} M_i = -23.27 + 38.41(i-z) - 11.11(i-z)^2  + 1.064(i-z)^3 \pm 0.41 \end{equation}
\begin{equation} M_i = 66.88 - 41.73(i-J) + 10.26(i-J)^2 - 0.7645(i-J)^3 \pm 0.33 \end{equation}

\noindent as well as the M$_J$, M$_H$, and M$_K$ versus optical spectral type relations from \citet{Looper2008b}. These distances and associated uncertainties (based on the error in magnitude and the scatter of the relations) are given in Table~\ref{tab:prop}. There is a relatively large spread in the distances estimates, likely due to the peculiar infrared colors of SDSS 1416+13. The weighted mean of the five estimates places SDSS 1416+13 at 8.0$\pm$1.6~pc (the uncertainty includes the error in each distance, as well as the standard deviation of the distances). Even if this source is binary (see Section 2.3) it is well within the 20~pc horizon used to determine the L dwarf luminosity function. 

\subsection{Near-Infrared Spectroscopy}
SDSS 1416+13 was observed with the 3m NASA Infrared Telescope Facility (IRTF) SpeX spectrograph \citep{Rayner2003} on 2009 June 28-29 (UT). Conditions were clear on both nights with seeing of 0$\farcs$5-0$\farcs$7 at $J$- band.  On June 28, we observed this source using the SpeX prism-dispersed mode, which provides 0.75--2.5~$\micron$ continuous spectroscopy with a resolution of $\approx 120$ for the 0$\farcs$5 slit employed (dispersion across the chip is 20--30~{\AA}~pixel$^{-1}$). The slit was aligned to the parallactic angle. Four exposures of 120~s each were obtained in an ABBA dither pattern along the slit, at an average airmass of 1.02. We observed the nearby A0 V star HD 121880 immediately after the SDSS 1416+13 observations at a similar airmass for telluric absorption correction and flux calibration, and internal quartz flat and Ar lamp exposures for pixel response and wavelength calibration. On June 29, we obtained spectra using the SpeX cross-dispersed mode, providing 0.8--2.4~$\micron$ spectroscopy in five orders at an average resolution of $\approx 1200$ for the 0$\farcs$5 slit.  Four exposures of 300s each were obtained in an ABBA dither pattern at an average air mass of 1.13. As with the prism observations, the slit was aligned to the parallactic
angle, and the A0 V star HD 124773 was subsequently observed, along with flat and arc exposures, for calibration.  All data were reduced using SpeXtool version 3.3 \citep{Vacca2003,Cushing2004} using standard settings (see \citealt{Cushing2005} and \citealt{Burgasser2008a} for details).

\begin{figure}
\includegraphics[width=0.97\linewidth]{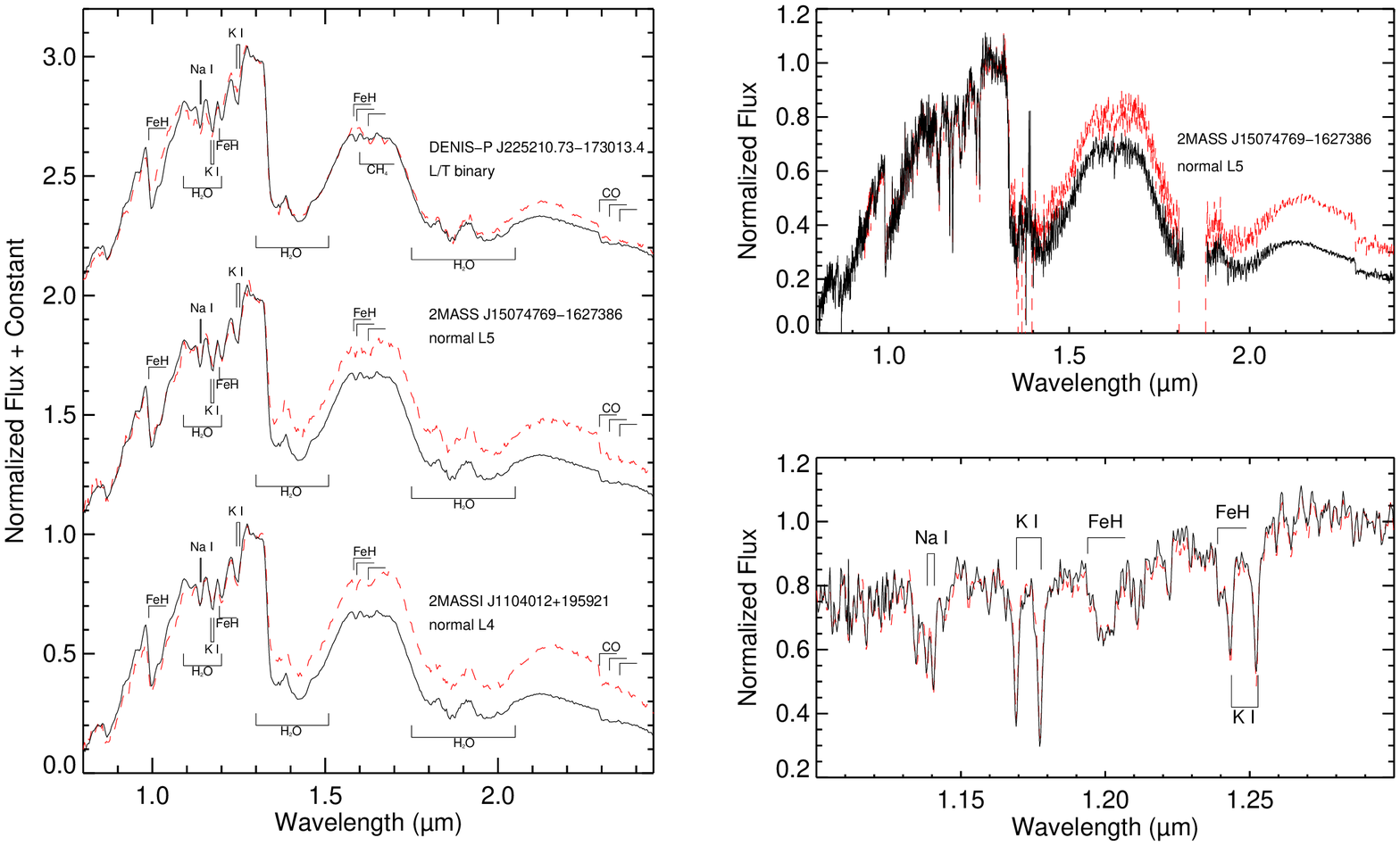} 
\caption{In the left panel, the IRTF SpeX prism-dispersed spectrum of SDSS~1416+13 (black) shown with spectra (red; from bottom to top) of 
2MASSI~J1104012+19592 \citep[L4;][]{Cruz2003,Burgasser2004a}, 
2MASSW~J15074769-1627386 \citep[L5;][]{Reid2000,Burgasser2008b}, and 
DENIS-P~J225210.73-173013.4 \citep[resolved L6/T2 binary;][]{Kendall2004,Reid2006a}.
The spectra are normalized at 1.3$\mu$m and major spectral features are labeled.
In the two right panels, the IRTF SpeX cross-dispersed spectrum of SDSS 1416+13 (black) compared with the spectrum of 2MASSW~J15074769-1627386 (from the IRTF spectral library \protect \footnote{Available at \url{http://irtfweb.ifa.hawaii.edu/$\sim$spex/IRTF\_Spectral\_Library}.}; Cushing et al.\ 2005). The top panel shows the entire spectrum, and the bottom panel shows the region from 1.1 to 1.3$\mu$m, which contains the Na I and K I doublets.} \label{fig:spex}
\end{figure}

In order to assign an infrared spectral type to SDSS 1416+13, we calculated several spectral indices from the literature, shown in Table~\ref{tab:IRt}. The resulting spectral types vary from L2 to L6 with a mean classification of L4$\pm$1.5. The variation is not surprising due to the peculiar colors of SDSS 1416+13, but the earlier near-infrared type compared to the optical classification is unusual, given that most blue L dwarfs are classified $\sim$2 subtypes later in the near-infrared \citep{Burgasser2008a}. The left panel of Figure~\ref{fig:spex} shows the comparison of the near-infrared spectrum of SDSS 1416+13 to L4 and L5 spectral standards. While the region between 1.1 and 1.3 $\mu$m (Na I and K I doublets) shows a good match to the L4 standard, the overall spectral slope of SDSS 1416+13 is too blue to match either of the spectral standards.

Some unusually blue L dwarfs have been resolved as L dwarf plus T dwarf binaries, the T dwarf component contributing weak absorption features and making the spectral slope bluer \citep{Burgasser2008b}. In Figure~\ref{fig:spex} we also compare the spectra of SDSS 1416+13 and DENIS-P J225210.73-173013.4, a resolved L/T binary \citep{Reid2006b} with estimated spectral types of L6 and T2. While the match of the spectral slope is better than for the L4 and L5 dwarfs, SDSS 1416+13 does not show any evidence of CH$_4$ in the $H$ band, which is typically present in composite L dwarf/T dwarf spectra. It is therefore unlikely that the blue color is due to unresolved binarity, although high resolution imaging to test this conjecture is warranted.

The cross-dispersed spectrum of SDSS 1416+13 is shown in the right panels of Figure~\ref{fig:spex}. Equivalent widths of the K I doublets at $\lambda\sim$  1.175 $\mu$m and $\lambda\sim$ 1.245 $\mu$m are given in Table~\ref{tab:EW}. The uncertainties were calculated by propagating the error in flux and the standard deviation of the continuum regions. SDSS 1416+13 has strong alkali absorption, with EW that are comparable to typical values for L4/L5 dwarfs measured at comparable resolution \citep{Cushing2005}.

\subsection{Kinematics}
To calculate the proper motion of SDSS 1416+13, we combined the SDSS and 2MASS coordinates with three positions from the SuperCOSMOS Sky Survey \citep{Hambly2001}. A fit to the five epochs gives $\mu_{\alpha}$ = 0.\arcsec0880 $\pm$ 0\arcsec.0028 yr$^{-1}$ and $\mu_{\delta}$ = 0\arcsec.1399 $\pm$ 0\arcsec.0013 yr$^{-1}$ with a total baseline of 49.07~years and uncertainties based on the scatter in the fit to right ascension and declination. The proper motion is remarkably small for an L dwarf within 10~pc of the Sun, and explains why this object has not been found in proper motion surveys to recover nearby ultracool dwarfs \citep[e.g.,][]{Deacon2005,Sheppard2009}.

The radial velocity of SDSS 1416+13 was calculated by measuring the line centers of the atomic lines\footnote{Line centers were obtained from the National Institute of Standards and Technology Web site.} Rb I (7802.41 and 7949.79 \AA), Na I (8185.51 and 8197.08 \AA) and Cs I (8523.47 and 8945.93 \AA) in the MagE spectrum.  The resulting line measurements were averaged to obtain a radial velocity of -42.2$\pm$5.1 km s$^{-1}$ (corrected to a solar frame of reference). The uncertainty reflects the scatter in the six line velocities. 

\textit{UVW} velocities and uncertainties were calculated using the method outlined in \citet{Johnson1987} and are given in Table~\ref{tab:prop}, corrected to the local standard of rest assuming $UVW_{\odot}$ = (-10,5,7) km s$^{-1}$ \citep{Dehnen1998}. We define positive $U$ as toward the Galactic center. SDSS 1416+13 meets the kinematic criteria for young-old disk membership, with $UV$ velocities just outside the young disk ellipsoid \citep{Eggen1969,Leggett1992}. An integration of the orbit of SDSS 1416+13 in the galactic potential \citep[after][]{Burgasser2009c} shows that it is nearly circular, with the object traveling up to 500~pc above and below the plane of the Galaxy during its orbital period. The kinematics of this source are consistent with its interpretation as a (possibly older) thin disk star, most likely with a stellar mass (M $>$ 0.077 M$_{\odot}$ for age $<$ 5 Gyr), a large surface gravity and possibly subsolar metallicity.

\section{Discussion}
\begin{figure}
\includegraphics[width=0.5\linewidth]{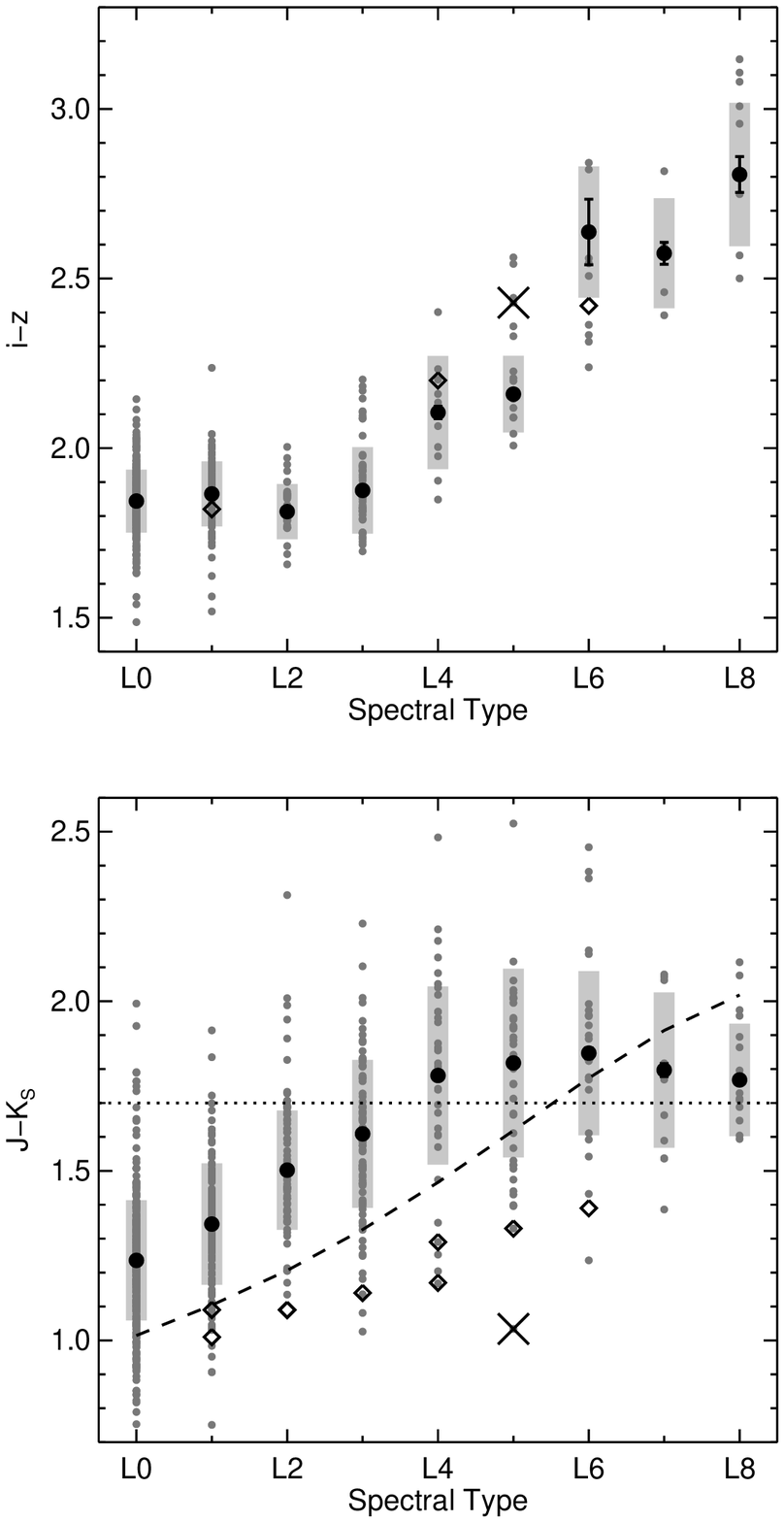} 
\caption{The $i-z$ (top panel) and $J-K_S$ (bottom panel) colors as a function of optical spectral type for L0 to L8 dwarfs. Half spectral types are rounded down. The median color (black circles) and standard deviation (grey bars) are shown with the colors of individual objects (grey circles). The colors of blue L dwarfs with optical spectral types are shown (diamonds) as well as the colors of SDSS 1416+13 (cross). The \textit{J$-$K$_S$} color selection criteria of  \citet[dotted line]{Kirkpatrick1999,Kirkpatrick2000} and \citet[dashed line]{Cruz2003,Cruz2007} and \citet{Reid2008} are shown in the bottom panel; the latter is a combination of the color-magnitude cut and the \citet{Cruz2003} spectral type/M$_J$ relation.} \label{fig:col}
\end{figure}

Figure~\ref{fig:col} shows the $i-z$ and \textit{J$-$K$_S$} colors of SDSS 1416+13 compared to the median colors of all known L dwarfs based on data from the Dwarf Archives\footnote{Available at \url{http://dwarfarchives.org}.} and \citet{Schmidt2009} according to optical spectral type; objects with only near-infrared spectral types were excluded. While SDSS 1416+13 is blue in \textit{J$-$K$_S$} color, its $i-z$ color is normal for its spectral type. This is typical of other blue L dwarfs, which are also shown in Figure~\ref{fig:col}. 

SDSS 1416+13 lies blueward of previous 2MASS color-color selection criteria. Initially, \citet{Kirkpatrick1999,Kirkpatrick2000} selected for dwarfs with $J-K_S \geq 1.7$, in order to increase the sample of the latest L dwarfs. To construct the luminosity function from the 20~pc sample, \citet{Cruz2003,Cruz2007} and \citet{Reid2008} created a color criteria in the $J$ vs. $J-K_S$ color magnitude diagram that included fainter sources if they had red colors to balance the need to obtain a complete sample with the heavy contamination by late-M dwarfs. SDSS 1416+13 falls far from both color cuts, and all blue L dwarfs fall to the blue side of the 2MASS color criteria used to assemble large samples of L dwarfs.

At a distance of 8.0$\pm$1.6~pc, SDSS 1416+13 will contribute to the 20~pc luminosity function  \citep{Cruz2007}. Its inclusion increases the number of L5 dwarfs within 20~pc from 4 to 5 -- a 20\% increase. In addition, with an absolute magnitude of M$_J$=13.7$\pm$0.1, it increases the number of dwarfs in the 13.5 - 14 mag bin from 6 to 7. While the inclusion of SDSS 1416+13 does not significantly change the shape of the luminosity function, the possible exclusion of other nearby blue L dwarfs (as yet undiscovered) may result in a substantial underestimate of the L dwarf space density. Current surveys do not sample the entire range of possible L dwarf colors and properties, but create a sample biased towards redder $J-K_S$ colors (Schmidt et al.\ 2009, submitted). The discovery of SDSS 1416+13 reinforces the need to probe the solar neighborhood for additional dwarfs using spectroscopy in addition to both optical and near-infrared photometry to mitigate selection biases.

\acknowledgments
This research has benefitted from the SpeX Prism Spectral Libraries, maintained by Adam Burgasser at http://www.browndwarfs.org/spexprism and from the M, L, and T dwarf compendium housed at DwarfArchives.org and maintained by Chris Gelino, Davy Kirkpatrick, and Adam Burgasser. 

This publication makes use of data from the Two Micron All Sky Survey, which is a joint project of the University of Massachusetts and the Infrared Processing and Analysis Center, and funded by the National Aeronautics and Space Administration and the National Science Foundation. 2MASS data were obtained from the NASA/IPAC Infrared Science Archive, which is operated by the Jet Propulsion Laboratory, California Institute of Technology, under contract with the National Aeronautics and Space Administration.

Funding for the SDSS and SDSS-II has been provided by the Alfred P. Sloan Foundation, the Participating Institutions, the National Science Foundation, the U.S. Department of Energy, the National Aeronautics and Space Administration, the Japanese Monbukagakusho, the Max Planck Society, and the Higher Education Funding Council for England. The SDSS Web Site is http://www.sdss.org/.

The SDSS is managed by the Astrophysical Research Consortium for the Participating Institutions. The Participating Institutions are the American Museum of Natural History, Astrophysical Institute Potsdam, University of Basel, University of Cambridge, Case Western Reserve University, University of Chicago, Drexel University, Fermilab, the Institute for Advanced Study, the Japan Participation Group, Johns Hopkins University, the Joint Institute for Nuclear Astrophysics, the Kavli Institute for Particle Astrophysics and Cosmology, the Korean Scientist Group, the Chinese Academy of Sciences (LAMOST), Los Alamos National Laboratory, the Max-Planck-Institute for Astronomy (MPIA), the Max-Planck-Institute for Astrophysics (MPA), New Mexico State University, Ohio State University, University of Pittsburgh, University of Portsmouth, Princeton University, the United States Naval Observatory, and the University of Washington.


\bibliographystyle{apj}

\begin{deluxetable}{ll} \tablewidth{0pt} \tabletypesize{\scriptsize}
\tablecaption{Properties of SDSS 1416+13\label{tab:prop}}
\tablehead{\colhead{Parameter}   & \colhead{SDSS 1416+13}} 
\startdata
SDSS Coordinates (J2000) & 14 16 24.09 +13 48 26.74\tablenotemark{1} \\
Optical ST & L5 \\
Near Infrared ST   & L4 $\pm$ 1.5 \\
\textit{r}  & 20.68 $\pm$ 0.04 \\
\textit{i}  & 18.37 $\pm$ 0.02 \\
\textit{z}  & 15.94 $\pm$ 0.02 \\
\textit{J}  & 13.15 $\pm$ 0.02 \\
\textit{H}  & 12.46 $\pm$ 0.03 \\
\textit{K$_s$}  & 12.11 $\pm$ 0.02 \\
\textit{(i-z)}  & 2.43 $\pm$ 0.02 \\
\textit{(J-K$_s$)}  & 1.03 $\pm$ 0.03 \\
Distance (M$_i$ vs. $i-z$) & 5.6 $\pm$ 1.1 pc \\
Distance (M$_i$ vs. $i-J$) & 5.2 $\pm$ 0.8 pc \\
Distance (M$_J$ vs. ST) & 7.7 $\pm$ 1.1 pc \\
Distance (M$_H$ vs. ST) & 9.3 $\pm$ 1.1 pc \\
Distance (M$_K$ vs. ST) & 11.0 $\pm$ 1.3 pc \\
Distance (mean) & 8.0 $\pm$ 1.6 pc \\
$\mu_{\alpha}$ & 0.0880 $\pm$ 0.0028 \arcsec yr$^{-1}$ \\
$\mu_{\delta}$ & 0.1399 $\pm$ 0.0013 \arcsec yr$^{-1}$ \\
V$_{\rm tan}$ & 6.31 $\pm$ 1.24 km s$^{-1}$ \\
Radial Velocity & -42.2 $\pm$ 5.1 km s$^{-1}$ \\
U & -7.9 $\pm$ 2.1 km s$^{-1}$ \\
V & 10.2 $\pm$ 1.2 km s$^{-1}$ \\
W & -31.4 $\pm$ 4.7 km s$^{-1}$ \\
\enddata
\tablenotetext{1}{Epoch 2003.41}
\end{deluxetable}

\begin{deluxetable}{llll} \tablewidth{0pt} \tabletypesize{\scriptsize}
\tablecaption{Near-Infrared Spectral Indices \label{tab:IRt}}
\tablehead{\colhead{Index}   & \colhead{Value} & \colhead{ST} & \colhead{Reference}} 
\startdata
K1                                  &   0.23  &   L3  &  1,2 \\
H$_2$O A                    &   0.67  &   L2  &  1 \\
H$_2$O B                    &   0.66  &   L4  &  1 \\
H$_2$O 1.5 $\mu$m &   1.60  &   L6  &  3 \\
CH$_4$ 2.2 $\mu$m &   1.02  &   L5  &  3 \\
H$_2$O(c)                   &   0.82  &   L4  & 4 \\
\enddata
\tablerefs{(1) \citet{Reid2001}; (2) \citet{Kobayashi1999}; (3) \citet{Geballe2002}; (4) \citet{Burgasser2008a}}
\end{deluxetable}

\begin{deluxetable}{lll} \tablewidth{0pt} \tabletypesize{\scriptsize}
\tablecaption{Equivalent Widths of Lines \label{tab:EW}}
\tablehead{\colhead{Species}   & \colhead{Wavelength} & \colhead{EW (\AA)}} 
\startdata
H$\alpha$  &   6563 \AA           &            $<$ 0.04 $\pm$ 1.3     \\
Li I                &   6807 \AA           &            $<$ 0.32 $\pm$ 0.4     \\
K I                &   1.169 $\mu$m  &  10.6 $\pm$ 0.3 \\
                    &    1.178 $\mu$m  &  13.6 $\pm$ 0.3 \\
                    &    1.244 $\mu$m  &  5.0   $\pm$ 0.5 \\
                    &    1.253 $\mu$m  &  8.7   $\pm$ 0.5 \\ 
\enddata
\end{deluxetable}

\end{document}